%% file: paper.tex
\documentclass[%
 reprint,
 amsmath,amssymb,
 aps,
]{revtex4-2}

\usepackage{graphicx}
\usepackage{dcolumn}
\usepackage{bm}
\usepackage[mathlines]{lineno}

\usepackage{units}
\usepackage{multirow}
\usepackage{physics}
\usepackage{csquotes}
\usepackage[normalem]{ulem}
\usepackage{todonotes}
\usepackage{amsmath,amsfonts,amssymb,amsthm}
\usepackage{hyperref}
\usepackage{orcidlink}

\newcommand{\Vcb}{V_\mathrm{cb}}
\newcommand{\cosv}{\cos \theta_\mathrm{V}}
\newcommand{\cosl}{\cos \theta_\mathrm{\ell}}

\newcommand{\cosvsq}{\cos^2 \theta_\mathrm{V}}

\newcommand{\sinvsq}{\sin^2 \theta_\mathrm{V}}
\newcommand{\sinlsq}{\sin^2 \theta_\mathrm{\ell}}

\newcommand{\mDst}{m_\mathrm{D^*}}
\newcommand{\GF}{G_\mathrm{F}}

\newcommand{\Bbar}{\,\overline{\!B}{}}

\begin{document}



\preprint{
Belle Preprint           2022-34,
KEK Preprint             2022-47
}

\title{Measurement of Angular Coefficients of  $\bar{B} \to D^* \ell \bar{\nu}_\ell$: Implications for $|V_{cb}|$ and Tests of Lepton Flavor Universality}

\input{authorsOrcid}

\begin{abstract}
We measure the complete set of angular coefficients $J_i$ for exclusive $\bar{B} \to D^* \ell \bar{\nu}_\ell$ decays ($\ell = e, \mu$). Our analysis uses the full $711\,\mathrm{fb}^{-1}$ Belle data set with hadronic tag-side reconstruction.
The results allow us to extract the form factors describing the $B \to D^*$ transition and the Cabibbo-Kobayashi-Maskawa matrix element $|V_{\rm cb}|$. Using recent lattice QCD calculations for the hadronic form factors, we find $|V_{\rm cb}| = (41.0 \pm 0.7) \times 10^3 $ using the BGL parameterization, compatible with determinations from inclusive semileptonic decays. We search for lepton flavor universality violation as a function of the hadronic recoil parameter $w$, and investigate the differences of the electron and muon angular distributions. We find no deviation from Standard Model expectations.
\end{abstract}

\maketitle


In this Letter, we present the first determination of the full set of angular coefficients describing the full differential decay rate of exclusive semileptonic $\bar{B} \to D^* \ell \bar{\nu}_\ell$ ($\ell = e, \mu$) decays. Our analysis uses the complete Belle data set, with an integrated luminosity of $\unit[711]{fb^{-1}}$ at the $\Upsilon(4S)$ resonance. The data set was recorded at the KEKB $e^+e^-$ collider~\cite{KEKB} by the Belle detector. Belle is a large-solid-angle magnetic spectrometer. A detailed description of its performance and subdetectors can be found in Ref.~\cite{Abashian:2000cg}.
We use hadronic tagging to reconstruct the accompanying $B$ meson. The measured angular coefficients allow us to determine the form factors that describe the non-perturbative dynamics describing the $B \to D^*$ transition and consequently, in conjunction with information from Lattice QCD (LQCD), to extract the Cabibbo-Kobayashi-Maskawa (CKM) matrix element $|V_{\rm cb}|$. The angular coefficients are also sensitive to beyond Standard Model (SM) effects and are used to test lepton flavor universality. Our measurement is based on the same dataset analyzed in a previous publication~\cite{Belle:2023bwv}, which focused on partial branching fractions in bins of the hadronic recoil parameter 

\begin{equation}
    w = \frac{m_B^2 + m_{D^*}^2 - q^2}{2m_B m_{D^*}} \,, 
\end{equation}
with the $B$ ($D^*$) mass $m_B$ ($m_{D^*}$) and the momentum-transfer squared to the lepton-neutrino system $q^2$,
and the decay angles $\theta_\ell$, $\theta_V$, and $\chi$. The decay angles are defined as follows: 
$\theta_\ell$ is the angle between the lepton and the direction opposite the $B$ meson in the virtual $W$-boson rest frame, 
$\theta_V$ is the angle between the $D$ meson and the direction opposite the $B$ meson in the $D^*$ rest frame, 
and $\chi$ is the angle between the two decay planes spanned by the $W^+ -\ell$ and $D^*-D$ systems in the $B$ meson rest frame.
The analysis strategy closely follows the methodology outlined in Ref.~\cite{Belle:2023bwv}, with modifications to facilitate the measurement of angular coefficients as a function of $w$. 

The four-dimensional differential decay rate for $\bar{B} \to D^* \ell \bar{\nu}_\ell$ can be expressed in terms of 12 functions $J_i = J_i(w)$, which only depend on $w$, as
\begin{widetext}
\begin{align}
     \frac{\dd \Gamma (\bar{B} \to D^* \ell \bar{\nu}_\ell)}{\dd w\, \dd\kern-0.14em\cosl\, \dd\kern-0.14em \cosv\, \dd \chi} =& \frac{2\GF^2 \eta_{\rm EW}^2  |\Vcb|^2 m_B^4 \mDst }{2\pi^4} \times \bigg( J_{1s} \sinvsq + J_{1c} \cosvsq \nonumber \\
     & + (J_{2s} \sinvsq + J_{2c} \cosvsq) \cos 2\theta_\ell + J_3 \sinvsq \sinlsq \cos 2\chi \nonumber \\
     & + J_{4} \sin 2\theta_V \sin 2\theta_\ell \cos\chi + J_{5} \sin 2\theta_V \sin \theta_\ell \cos\chi + (J_{6s} \sinvsq + J_{6c} \cosvsq) \cos \theta_\ell \nonumber \\ 
     & + J_{7} \sin 2\theta_V \sin \theta_\ell \sin\chi + J_{8} \sin 2\theta_V \sin 2\theta_\ell \sin\chi + J_{9} \sinvsq \sinlsq \sin 2\chi \bigg) \,.
\end{align}
\end{widetext}
The expression depends on Fermi's coupling constant $G_{\rm F}$, the electroweak correction $\eta_{\rm EW}$~\cite{Sirlin:1981ie}, the CKM matrix element $V_{\rm cb}$, and the masses of $B$ ($m_B$) and $D^*$ ($m_{D^*}$) mesons. 

We determine the angular coefficients in bins of $w$, $\bar{J}_i = \int_{\Delta w} J_i(w) \dd w$, from experimental data with the definition from Ref.~\cite{Bernlochner:2014ova}:
\begin{align}
    \bar{J}_i = \frac{1}{N_i} \sum_{j=1}^{8} \sum_{k,l=1}^{4} \eta_{i,j}^\chi \eta_{i,k}^{\theta_\ell} \eta_{i,l}^{\theta_V} R_{jkl} \,.
    \label{eq:yield_to_J}
\end{align}
The normalization factor $N_i$ stems from trigonometric integrals. The angles $\theta_\ell$, $\theta_V$, and $\chi$ are divided into bins of size $\pi/4$. The weight factors $\eta_{i,n}^\alpha$ with $\alpha \in \{\chi, \theta_\ell, \theta_V\}$ are given in Ref.~\cite{Bernlochner:2014ova} and the product of these factors define a specific phase-space bin where signal is extracted. The factor $R_{jkl}$ represents the partial rate in the corresponding phase-space bin $jkl$. 
We combine phase-space bins with identical products of the weights $\eta_{i,n}^\alpha$ during signal extraction, resulting in yields of total 36 merged bins to obtain 12 $J_i$ coefficients using Eq.~(\ref{eq:yield_to_J}) in each bin of $w$.
In the limit of massless charged leptons, the angular coefficient $J_{6c}$ vanishes. Furthermore, the angular coefficients $J_{7}$, $J_{8}$, and $J_{9}$ are zero within the SM of particle physics, only contributing to scenarios involving new physics.


We reconstruct two $B$ meson candidates, a tag $B$ and a signal $B$. Signal $B$ meson candidates are reconstructed as follows: 
We consider both charged and neutral $B$ mesons with the decay chains $\Bbar{}^0 \to D^{*+} \ell \overline{\nu}_\ell$, $D^{*+} \to D^0 \pi^+$, $D^{*+} \to D^+ \pi^0$, and $B^- \to D^{*0} \ell \overline{\nu}\ell$ with $D^{*0} \to D^0\pi^0$ \footnote{Charge conjugation is implicit throughout this Letter.}.

To select charged tracks, we apply the following criteria: $dr < \unit[2]{cm}$ and $|dz| < \unit[4]{cm}$, where $dr$ is the impact parameter perpendicular to the beam-axis and with respect to the interaction point and $dz$ is the $z$ coordinate along the beam-axis of the impact parameter. Tracks are also required to have transverse momenta $p_T > \unit[0.1]{GeV/c}$. In addition, we utilize particle identification subsystems to identify electrons, muons, charged pions, kaons, and protons. Electron (muon) tracks are required to have momenta in the lab frame $p^\mathrm{Lab} > \unit[0.3]{GeV/c}$ ($p^\mathrm{Lab} > \unit[0.6]{GeV/c}$). The momenta of particles identified as electrons are corrected for bremsstrahlung by including photons within a $2^\circ$ cone defined around the electron momentum at the point of closest approach to the interaction point (IP).

Photon selection criteria are based on their energies: $E_\gamma > \unit[100]{MeV}$ for the forward endcap ($12^\circ < \theta < 31^\circ$), $\unit[150]{MeV}$ for the backward endcap ($132^\circ < \theta < 157^\circ$), and $\unit[50]{MeV}$ for the barrel region ($32^\circ < \theta < 129^\circ$) of the calorimeter.
$\pi^0$ candidates are formed from pairs of photons with invariant mass within the range of $\unit[104]{MeV/c^2}$ to $\unit[165]{MeV/c^2}$. The difference between the reconstructed $\pi^0$ mass and the nominal mass ($m_{\pi^0} = 135 \, \mathrm{MeV/c^2}$~\cite{pdg:2022}) must be smaller than three times the estimated mass resolution.

$K_\mathrm{S}^0$ mesons are reconstructed from oppositely charged track pairs within a reconstructed invariant mass window of $\unit[398]{MeV/c^2}$ to $\unit[598]{MeV/c^2}$ and selected with a multivate method. For details on the multivariate method used, see Ref.~\cite{Belle:2018xst}. The reconstructed $K_\mathrm{S}^0$ mass has to differ from the nominal value ($m_{K_\mathrm{S}^0} = 498 \, \mathrm{MeV/c^2}$~\cite{pdg:2022}) by less than $3\sigma$ of the estimated mass resolution. 

We reconstruct the following decays of the $D$ mesons:
$D^+ \to K^- \pi^+ \pi^+$,
$D^+ \to K^- \pi^+ \pi^+ \pi^0$,
$D^+ \to K^- \pi^+ \pi^+ \pi^+ \pi^-$,
$D^+ \to K_\mathrm{S}^0 \pi^+$,
$D^+ \to K_\mathrm{S}^0 \pi^+ \pi^0$,
$D^+ \to K_\mathrm{S}^0 \pi^+ \pi^+ \pi^-$,
$D^+ \to K_\mathrm{S}^0 K^+$,
$D^+ \to K^+ K^- \pi^+$,
$D^0 \to K^- \pi^+$,
$D^0 \to K^- \pi^+ \pi^0$,
$D^0 \to K^- \pi^+ \pi^+ \pi^-$,
$D^0 \to K^- \pi^+ \pi^+ \pi^- \pi^0$,
$D^0 \to K_\mathrm{S}^0 \pi^0$,
$D^0 \to K_\mathrm{S}^0 \pi^+ \pi^-$,
$D^0 \to K_\mathrm{S}^0 \pi^+ \pi^- \pi^0$, and
$D^0 \to K^- K^+$.
We apply a decay-channel-optimized mass window selection to the $D$ meson candidates. 
The $\pi^0$ daughters from $D$ meson candidates are required to have center-of-mass momenta $p_{\pi^0}^\mathrm{CMS} > \unit[0.2]{GeV/c}$, except for the decay $D^0 \to K^- \pi^+ \pi^+ \pi^- \pi^0$ where this criterion is not applied. To reduce combinatorial background, the reconstructed $D$ mesons within an event are ranked based on the absolute difference between the reconstructed mass and the nominal mass ($m_{D^+} = 1.870 \, \mathrm{GeV/c^2}$, $m_{D^0} = 1.865 \, \mathrm{GeV/c^2}$~\cite{pdg:2022}), and up to ten candidates with the smallest mass difference are selected.

$D^*$ mesons are reconstructed in three decay channels: $D^{*0}\to D^0\pi^0_\mathrm{slow}$, $D^{*+}\to D^+\pi^0_\mathrm{slow}$, and $D^{*+}\to D^0\pi^+_\mathrm{slow}$. Charged slow pions must have a center-of-mass momentum below $\unit[0.4]{GeV/c}$, and the mass difference between the reconstructed masses $M_X$ of the $D^*$ and $D$ candidates $\Delta M = M_{D^*}- M_D$ has to be smaller than $\unit[0.155]{GeV/c^2}$ ($\unit[0.160]{GeV/c^2}$) for $D^{*+}$ ($D^{*0}$) mesons.

Signal-$B$ meson candidates are reconstructed by combining selected $D^*$ candidates and a lepton candidate. 
The loose selection $\unit[1]{GeV/c^2} < M_{D^*\ell} < \unit[6]{GeV/c^2}$ is applied to reduce combinatorial background.

We perform global-decay-chain vertex fitting using \texttt{TreeFitter} \cite{Belle-IIanalysissoftwareGroup:2019dlq}. Events that cannot be successfully fitted are rejected.

Tag-$B$ meson candidates are reconstructed using the Full Event Interpretation (FEI)~\cite{Keck:2018lcd} algorithm. The algorithm fully reconstructs $B$ mesons in 36 $B$-decay modes. Selection criteria include the requirement $ M_\mathrm{bc}^{\mathrm{tag}} = \sqrt{s/2 - \vec p_{\mathrm{tag}}^{\,2}} > \unit[5.27]{GeV/c^2}$ , $\Delta E_{\mathrm{tag}} = E_{\mathrm{tag}} - \sqrt{s}/2  \in [-0.15, 0.10]\, \mathrm{GeV}$, where $p_{\mathrm{tag}} = (E_{\mathrm{tag}} , \vec p_{\mathrm{tag}})$ is the 4-momentum of the $B_\mathrm{tag}$ meson in the center-of-mass frame, and the probability of the classifier for the tag-candidate $\mathcal{P}_\mathrm{FEI} > 10^{-3}$. We form an $\Upsilon(4S)$ candidate from combinations of tag- and signal-$B$ mesons, requiring that there are no additional charged particles in the event. The reconstructed invariant mass of the $\Upsilon(4S)$ candidate is limited to $M_{\Upsilon(4S)} \in [7.0, 13.0], \mathrm{GeV/c^2}$.

Non-resonant $e^+e^-$ interactions are suppressed using event shape variables: The magnitude of the thrust of the full event~\cite{BaBar:2014omp}, the angle between the thrust axis of the tag-$B$ and the beamline, the angle between the thrust axes of the two $B$ mesons~\cite{BaBar:2014omp}, the reduced Fox-Wolfram moment $R_2$~\cite{BaBar:2014omp}, the modified Fox-Wolfram moments~\cite{SFW}, and the CLEO Cones~\cite{cleocones}. These variables are combined using a multivariate classifier based on gradient boosted decision trees~\cite{Keck:2017gsv}.

The average number of $\Upsilon(4S)$ candidates is 4.3. In events with more than one candidate we retain only the candidate with the lowest $E_\mathrm{ECL}$, the sum of unassigned photon clusters in the full event reconstruction. If multiple candidates remain, the one with the smallest $|\Delta E_\mathrm{tag}|$ is chosen. If a conclusive selection cannot be made, a random candidate is selected.


The angular coefficients are extracted in four bins of $w$. In each $w$ bin, we determine the signal yield in bins of the decay angles $\theta_\ell$, $\theta_V$, and $\chi$, defined in Equation~\ref{eq:yield_to_J}. The signal yield in the bins of $36$ angles $\times$ $4$ $w$ bins $\times$ $4$ decay modes is determined using the $M_\mathrm{miss}^2$ distribution, where $M_\mathrm{miss}^2 = \left( p_{e^+e^-} - p_\mathrm{tag} - p_{D^*} - p_\ell \right)^2$. 
A binned maximum likelihood fit is performed using Monte-Carlo (MC) simulations to determine template shapes for signal and background events. Nuisance parameters account for systematic effects on the shapes of the templates in the fit. The $M_\mathrm{miss}^2$ distribution is binned into five bins to reduce dependence on the modelling of the $M_\mathrm{miss}^2$ resolution. The statistical correlation of the partial rates in different phase-space regions projected onto $M_\mathrm{miss}^2$ is determined by bootstrapping.

The determined signal yields are transformed into partial rates $\Delta \Gamma /\Delta x$ through a process of unfolding, utilizing the matrix inversion method, and subsequent correction for efficiency. A more detailed description of the individual steps can be found in Ref.~\cite{Belle:2023bwv}. Unfolding is necessary to correct for resolution effects causing migration of signal events into different regions of phase-space. Systematic uncertainties in the migration matrix and efficiency correction are accounted for by reweighting the simulated events. The most significant source of systematic uncertainty is from the limited available MC sample size used to derive migration matrices and efficiency corrections. Smaller uncertainties arise from $D$ decay branching fractions~\cite{pdg:2022}, assumptions about $B\to D^*\ell \bar{\nu}_\ell$ form factors~\cite{Belle:2018ezy}, the impact of the FEI on the measured shapes, lepton identification efficiency, and the efficiencies for reconstructing tracks, neutral pions, slow pions, and $K_{\rm S}^0$ mesons.
 
Subsequently, the angular coefficients $\bar{J}_i$ are calculated from the $\Delta \Gamma /\Delta x$ using Eq.~(\ref{eq:yield_to_J}). Due to the challenges in calibrating the absolute efficiency of the FEI, we quote normalized angular coefficients $\hat{J}_i = J_i / \mathcal{N}$, with $\mathcal{N}=\frac{8}{9}\pi \sum_{i=0}^{w \rm bins} 3 \bar{J}_{1c}^{(i)} + 6 \bar{J}_{1s}^{(i)} - \bar{J}_{2c}^{(i)} - 2 \bar{J}_{2s}^{(i)}$ summing over the four $w$ bins. 

\begin{figure*}
    \centering
    \includegraphics[width=\linewidth]{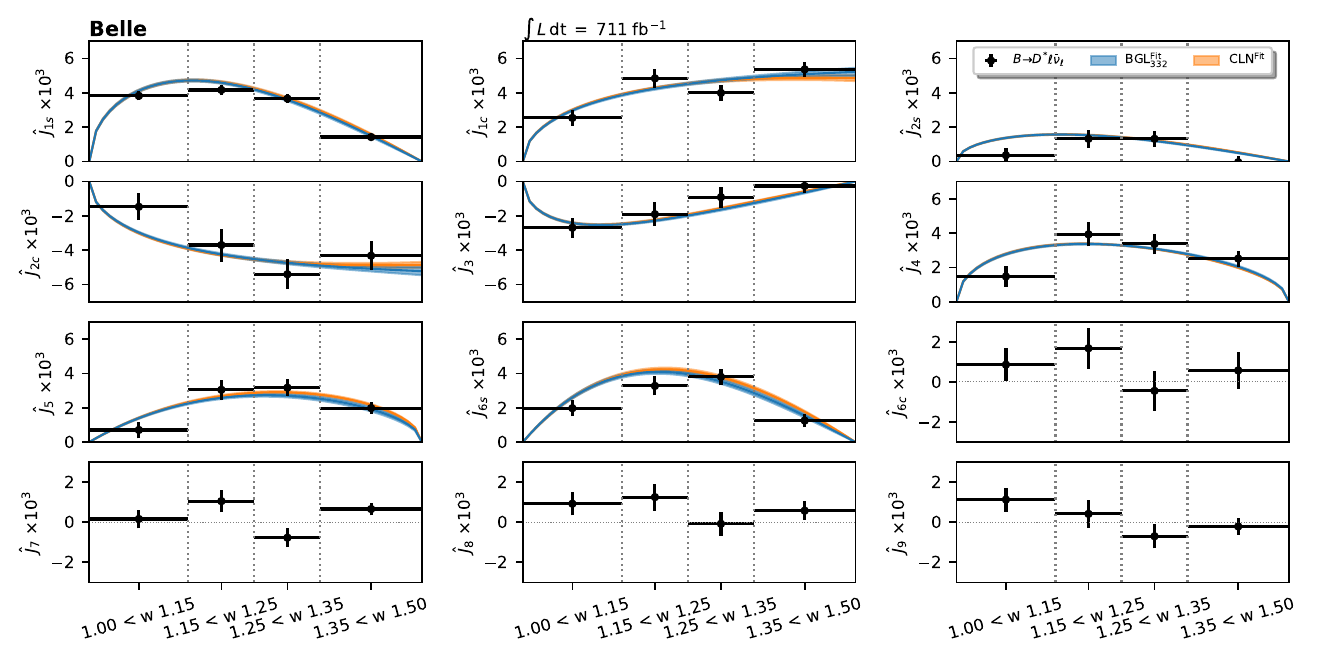}
    \caption{The data points correspond to the averaged central values of the four measured normalized angular coefficients described in the text, with the uncertainties including statistical and systematic uncertainties. The vertical dotted lines indicate the binning in $w$. The blue (orange) curves correspond to the BGL$_{332}$ (CLN) fit described in the text, with the $1\sigma$ uncertainty band. The angular coefficients $J_{6c}$, $J_7$, $J_8$, $J_9$ are not fitted, and expected to be zero in the SM.}
    \label{fig:angular-coefficients}
\end{figure*}

The analysis is validated using Asimov data and pseudo-experiments, for which we see no biases in central values or uncertainties. The measured normalized angular coefficients $\hat{J}_i$ as a function of $w$ are displayed in Fig.~\ref{fig:angular-coefficients}.

We calculate the average of the $\hat{J}_i$ determined in the four decay modes, taking into account the correlations and uncertainties. We fit the CLN~\cite{Caprini:1997mu} and BGL~\cite{Boyd:1995sq,Boyd:1997kz} form factor parameterizations to the averaged central values of the normalized measured angular coefficients within the SM (i.e.\,$J_{7,8,9}=0$) and neglecting electron and muon masses (so that $J_{6c}=0$). Beyond zero-recoil lattice calculations by the MILC~\cite{FermilabLattice:2021cdg}, HPQCD~\cite{Harrison:2023dzh}, and JLQCD~\cite{Aoki:2023qpa} groups are included in the fit to constrain the form factors. 
To obtain $|V_{\rm cb}|$ from the normalized angular coefficients, we use the absolute branching fraction of $\mathcal{B}(\bar{B}^0\to D^{*+} \ell \bar{\nu}_\ell) = (5.03 \pm 0.10)\%$. This value is obtained from the branching fractions $\mathcal{B}(B^-\to D^{*0} \ell \bar{\nu}_\ell) = (5.58 \pm 0.22)\%$, $\mathcal{B}(\bar{B}^0\to D^{*+} \ell \bar{\nu}_\ell) = (4.97 \pm 0.12)\%$~\cite{Amhis:2022mac} and the lifetimes $\tau_{\bar{B}^0} = 1.520 \, \mathrm{ps}$ and $\tau_{B^-} = 1.638 \, \mathrm{ps}$~\cite{pdg:2022}, assuming isospin symmetry.

The fit is performed by minimizing the $\chi^2$ function defined as
\begin{align}
 \chi^2 &= \left(\hat{J}_i^{\rm m} - \hat{J}_i^{\rm p}(\vec x)\right) C^{-1}_\mathrm{exp} \left(\hat{J}_i^{\rm m} - \hat{J}_i^{\rm p}(\vec x)\right)^T \nonumber \\
   & + (\Gamma^\mathrm{m} - \Gamma^{\rm p}(V_{\rm cb}, \vec{x}))^2 / \sigma(\Gamma^\mathrm{m})^2 \nonumber \\
   & + (h_{X}^{\rm LQCD} - h_{X}^{\rm p}(\vec{x})) C^{-1}_\mathrm{LQCD} (h_{X}^{\rm LQCD} - h_{X}^{\rm p}(\vec{x})) \,,
\end{align}
which compares the measured (predicted) normalized angular coefficients $\Delta \hat{J}_i^\mathrm{m (p)}$ in bins of $w$. The quantity $\Gamma^\mathrm{m(p)}$ represents the predicted (externally measured) absolute rate. The predicted rate is a function of the form factor coefficients $\vec{x}$ and $|V_{\rm cb}|$, taking $m_B = \unit[5.28]{GeV/c^2}$, $m_{D^*} = \unit[2.01]{GeV/c^2}$, and $m_e = m_\mu = 0$. The parameters $h_{A_1}$, $R_1$, $R_2$ are used for the predicted (LQCD) form factors $h_{X}^{\rm p(LQCD)}$ and the form factors are evaluated at the $w$ values provided by the lattice QCD predictions~\cite{FermilabLattice:2021cdg,Harrison:2023dzh,Aoki:2023qpa}. The three available lattice QCD predictions are used simultaneously.
The covariance matrix $C_\mathrm{exp}$ ($C_\mathrm{LQCD}$) corresponds to the experimental (lattice) data. We perform a nested hypothesis test~\cite{Bernlochner:2019ldg} to determine the number of BGL coefficients required to describe the data, resulting in the choice $a=3$, $b=3$, $c=2$, which are the number of free expansion coefficients for the BGL $g$, $f$, $\mathcal{F}$ form factors, respectively. The $p$-values for the BGL$_{332}$ and CLN fits are 0.75 and 0.39, respectively. The fitted angular coefficients are shown in Fig.~\ref{fig:angular-coefficients}. The resulting form factors, together with the lattice data used in the fit, are shown in Fig.~\ref{fig:form-factors}~\footnote{Tabulated results and further details can be found in the supplemental material.}. We find consistent values for the CKM matrix element $|V_{\rm cb}|$ for both form factor parameterizations:
\begin{align}
    |V_{\rm cb}| &= (41.0 \pm 0.3 \pm 0.4 \pm 0.5) \times 10^{-3}\, ({\rm BGL}_{332}) \nonumber \,, \\
    |V_{\rm cb}| &= (40.9 \pm 0.3 \pm 0.4 \pm 0.4) \times 10^{-3}\, ({\rm CLN})       \nonumber \,,
\end{align}
where the first uncertainty is from the measured data, the second uncertainty is from the external branching fraction, and the third uncertainty is from the LQCD inputs.

\begin{figure*}
    \centering
    \includegraphics[width=\linewidth]{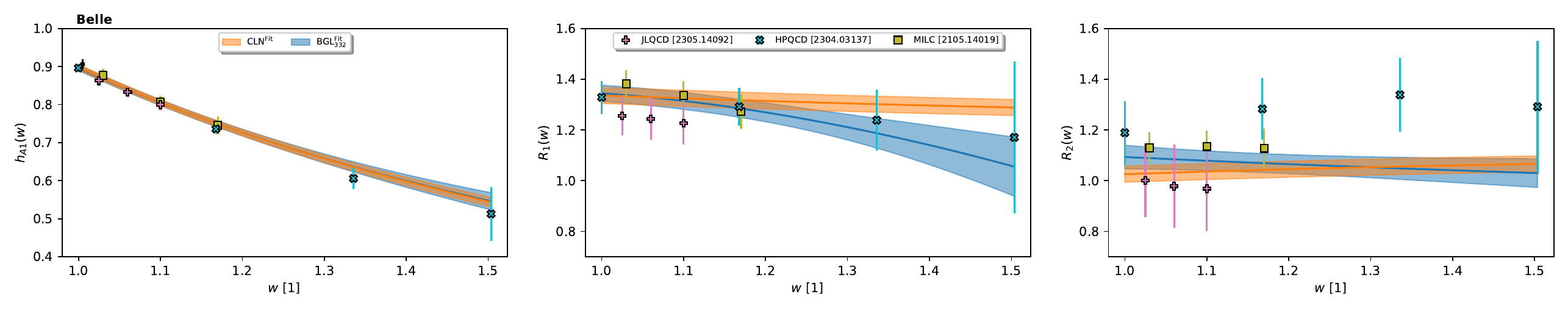}
    \caption{The blue (orange) band corresponds to our form factor fits using the BGL$_{332}$ (CLN) parameterizations with the beyond zero-recoil lattice predictions by (olive square) MILC~\cite{FermilabLattice:2021cdg}, (cyan cross) HPQCD~\cite{Harrison:2023dzh}, and (pink plus) JLQCD~\cite{Aoki:2023qpa} as input.}
    \label{fig:form-factors}
\end{figure*}

\begin{figure*}
    \centering
    \includegraphics[width=\linewidth]{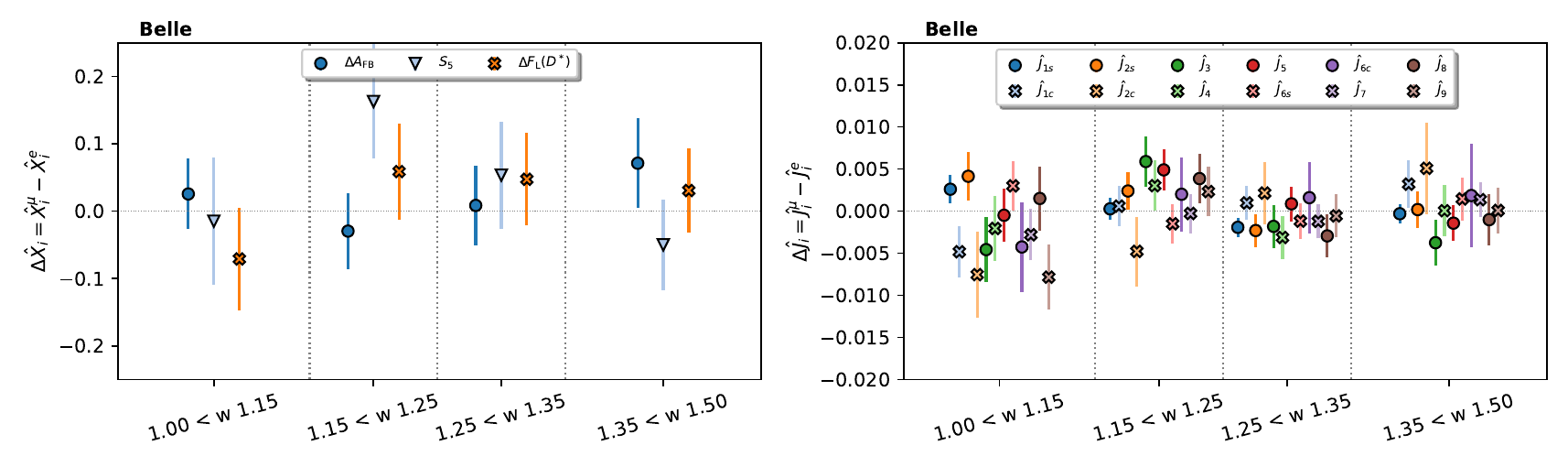}
    \caption{The results of the lepton flavor universality test described in the text. The $\Delta X = X^\mu - X^e$ are the observables testing the lepton flavor universal by calculating the difference between the decays with muons and electrons. $\Delta S_5 \propto \Delta J_5$ is shown since a new physics signal should show up simultaneously in $\Delta A_{FB}$ and $\Delta S_5$. These asymmetries are expected to be approximately zero in the SM.}
    \label{fig:observables}
\end{figure*}

The lepton forward-backward asymmetry $A_{\rm FB}$ and the $D^{*}$ longitudinal polarization fraction $F_L(D^*)$ are straightforwardly calculated~\footnote{For more details, including necessary relations, refer to the supplemental material} from the measured angular coefficients within their corresponding $w$ bins. The $S_i$ observables in Ref.~\cite{Bobeth:2021lya} are directly proportional to the angular coefficients $S_i \propto \hat{J}_i$ and are discussed further in the supplemental material. 
These observables can be used to test lepton flavor universality between electrons and muons via, e.g. $\Delta A_{\rm FB} = A_{\rm FB}^\mu - A_{\rm FB}^e$ to search for new physics effects. We observe no significant deviation from the SM expectation and quantify the compatibility of each observable with the SM expectation in Table~\ref{tab:lfu-test}. The corresponding lepton flavor universality observables are displayed in Fig.~\ref{fig:observables}.
\begin{table}[]
    \centering
    \caption{Compatibility of the lepton flavor universality observables with the SM expectation. The $\Delta X = X^\mu - X^e$ are the observables testing the lepton flavor universal by calculating the difference between the decays with muons and electrons.}
    \label{tab:lfu-test}
    \begin{tabular}{lcc}
    \hline
    \hline
    Observable & $\chi^2$ / ndf & $p$-value \\
    \hline
$\Delta A_{\rm FB}$ & 1.7 / 4 & 0.79 \\
$\Delta F_{\rm L}(D^*)$ & 2.3 / 4 & 0.67 \\
    \hline
$\Delta \hat{J}_{1s}$ & 5.3 / 4 & 0.26 \\
$\Delta \hat{J}_{1c}$ & 4.2 / 4 & 0.38 \\
$\Delta \hat{J}_{2s}$ & 4.6 / 4 & 0.33 \\
$\Delta \hat{J}_{2c}$ & 5.0 / 4 & 0.28 \\
$\Delta \hat{J}_{3}$ & 7.4 / 4 & 0.12 \\
$\Delta \hat{J}_{4}$ & 2.5 / 4 & 0.64 \\
$\Delta \hat{J}_{5}$ & 4.8 / 4 & 0.31 \\
$\Delta \hat{J}_{6s}$ & 2.1 / 4 & 0.72 \\
$\Delta \hat{J}_{6c}$ & 1.1 / 4 & 0.89 \\
$\Delta \hat{J}_{7}$ & 1.6 / 4 & 0.81 \\
$\Delta \hat{J}_{8}$ & 3.3 / 4 & 0.51 \\
$\Delta \hat{J}_{9}$ & 4.6 / 4 & 0.33 \\
\hline
$\Delta \hat{J}_i$ & 41 / 48 & 0.76 \\
\hline 
\hline
    \end{tabular}
\end{table}

In summary, we present the first complete measurement of the angular coefficients $\hat{J}_i$ in bins of $w$ describing the full differential decay distribution of $\bar{B} \to D^* \ell \bar{\nu}_\ell$ ($\ell = e, \mu$), probing both $\bar{B}^0$ and $B^-$ modes. In total, we measure the partial rates in $4 \times 144$ distinct phase-space regions to extract the $4\times12$ $\hat{J}_i$ coefficients, with full statistical and systematic correlations, allowing the simultaneous analysis of all measured angular coefficients. The measured coefficients encode the full angular information of the $\bar{B} \to D^* \ell \bar{\nu}_\ell$ decay, providing a more comprehensive set of observables than the one-dimensional partial rates of $w$, $\cos \theta_\ell$, $\cos \theta_V$, and $\chi$ measured in Refs.~\cite{Belle:2023bwv,Belle:2018ezy}.

\begin{figure}
    \centering
    \includegraphics[width=\linewidth]{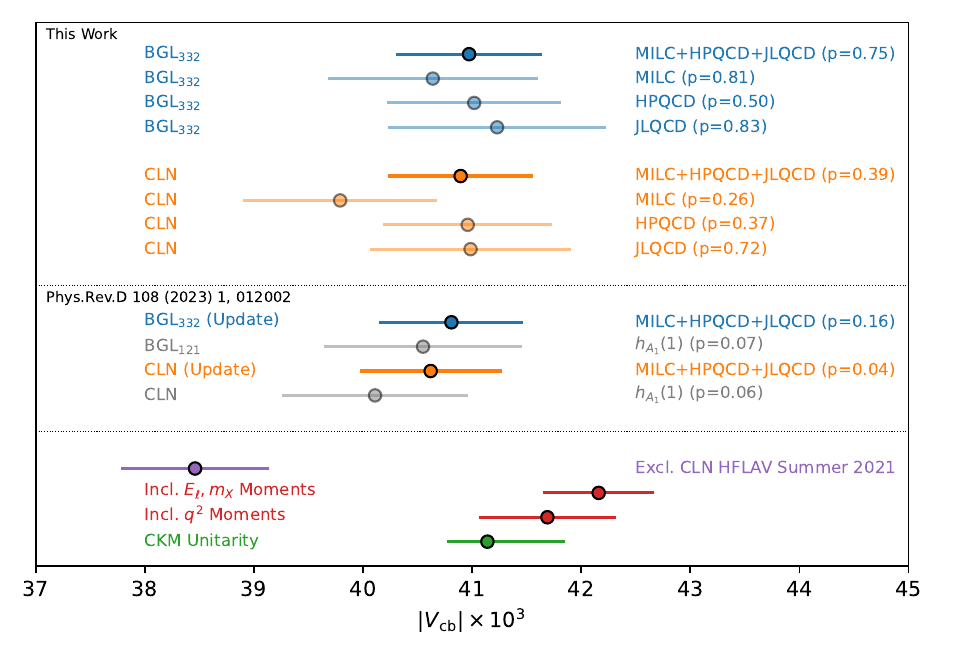}
    \caption{The results of the $|V_{\rm cb}|$ determination described in the text with other previous determinations. The top section shows the results of the analysis presented in this manuscript. The middle section shows the results in Ref.~\cite{Belle:2023bwv}, where we have updated the fit with beyond zero-recoil lattice data. The bottom section shows the HVLAV~\cite{HFLAV:2022pwe} world average of $|V_{\rm cb}|$, the $|V_{\rm cb}|$ determinations from inclusive decays~\cite{Bordone:2021oof,Bernlochner:2022ucr}, and $|V_{\rm cb}|$ determination from CKM unitarity.
    The BGL and CLN labels indicate the form factor paramterization used to determine $|V_{\rm cb}|$. The lattice QCD inputs are MILC~\cite{FermilabLattice:2021cdg}, HPQCD~\cite{Harrison:2023dzh}, JLQCD~\cite{Aoki:2023qpa}. Numbers in parentheses show goodness-of-fit $p$-values for the corresponding fits.
    }
    \label{fig:vcb-overview}
\end{figure}

The measured angular coefficients are analyzed to determine $|V_{\rm cb}|$ using the beyond zero-recoil lattice calculations by the MILC, HPQCD, and JLQCD collaborations and the world average of the $\bar{B} \to D^* \ell \bar{\nu}_\ell$ branching fraction and $B$-meson lifetimes. We find $|V_{\rm cb}| = (41.0 \pm 0.7) \times 10^{-3}$ in the BGL parameterization. The origin of the upward shift of $|V_{\rm cb}|$ with respect to Ref.~\cite{Belle:2023bwv} is caused by the shift of $\mathcal{F}(1) = 0.895\pm 0.007$ in the average of the new lattice results, and the smaller slope of the form factor compared to previous results. The resulting $p$-value of the fit is 75\% and the value of $|V_{\rm cb}|$ is in agreement with the fit of the one-dimensional partial rates determined from the same data set. The obtained values of $|V_{\rm cb}|$ are compatible with the determinations using the CLN parameterization. These results are also in agreement with the two currently most precise determinations of $|V_{\rm cb}|$ from inclusive $B \to X_c \ell \bar{\nu}_\ell$ measurements relying on heavy quark effective theory~\cite{Bordone:2021oof,Bernlochner:2022ucr}. 
Our results are in agreement with those determined from partial rates~\cite{Belle:2023bwv}, which uses the same dataset. A summary of our measurement of $|V_{\rm cb}|$, together with other determinations, is shown in Fig.~\ref{fig:vcb-overview}. 

The measured angular coefficients are tested for lepton flavor universality violation, and no deviation from the SM expectation is observed. The numerical values, and full covariance matrices of the measured observables have been made available on \texttt{HEPData}.

\acknowledgments

This work, based on data collected using the Belle detector, which was
operated until June 2010, was supported by 
the Ministry of Education, Culture, Sports, Science, and
Technology (MEXT) of Japan, the Japan Society for the 
Promotion of Science (JSPS), and the Tau-Lepton Physics 
Research Center of Nagoya University; 
the Australian Research Council including grants
DP210101900, 
DP210102831, 
DE220100462, 
LE210100098, 
LE230100085; 
Austrian Federal Ministry of Education, Science and Research (FWF) and
FWF Austrian Science Fund No.~P~31361-N36;
National Key R\&D Program of China under Contract No.~2022YFA1601903,
National Natural Science Foundation of China and research grants
No.~11575017,
No.~11761141009, 
No.~11705209, 
No.~11975076, 
No.~12135005, 
No.~12150004, 
No.~12161141008, 
and
No.~12175041, 
and Shandong Provincial Natural Science Foundation Project ZR2022JQ02;
the Czech Science Foundation Grant No. 22-18469S;
Horizon 2020 ERC Advanced Grant No.~884719 and ERC Starting Grant No.~947006 ``InterLeptons'' (European Union);
the Carl Zeiss Foundation, the Deutsche Forschungsgemeinschaft, the
Excellence Cluster Universe, and the VolkswagenStiftung;
the Department of Atomic Energy (Project Identification No. RTI 4002), the Department of Science and Technology of India,
and the UPES (India) SEED finding programs Nos. UPES/R\&D-SEED-INFRA/17052023/01 and UPES/R\&D-SOE/20062022/06; 
the Istituto Nazionale di Fisica Nucleare of Italy; 
National Research Foundation (NRF) of Korea Grant
Nos.~2016R1\-D1A1B\-02012900, 2018R1\-A2B\-3003643,
2018R1\-A6A1A\-06024970, RS\-2022\-00197659,
2019R1\-I1A3A\-01058933, 2021R1\-A6A1A\-03043957,
2021R1\-F1A\-1060423, 2021R1\-F1A\-1064008, 2022R1\-A2C\-1003993;
Radiation Science Research Institute, Foreign Large-size Research Facility Application Supporting project, the Global Science Experimental Data Hub Center of the Korea Institute of Science and Technology Information and KREONET/GLORIAD;
the Polish Ministry of Science and Higher Education and 
the National Science Center;
the Ministry of Science and Higher Education of the Russian Federation, Agreement 14.W03.31.0026, 
and the HSE University Basic Research Program, Moscow; 
University of Tabuk research grants
S-1440-0321, S-0256-1438, and S-0280-1439 (Saudi Arabia);
the Slovenian Research Agency Grant Nos. J1-9124 and P1-0135;
Ikerbasque, Basque Foundation for Science, and the State Agency for Research
of the Spanish Ministry of Science and Innovation through Grant No. PID2022-136510NB-C33 (Spain);
the Swiss National Science Foundation; 
the Ministry of Education and the National Science and Technology Council of Taiwan;
and the United States Department of Energy and the National Science Foundation.
These acknowledgements are not to be interpreted as an endorsement of any
statement made by any of our institutes, funding agencies, governments, or
their representatives.

We thank the KEKB group for the excellent operation of the
accelerator; the KEK cryogenics group for the efficient
operation of the solenoid; and the KEK computer group and the Pacific Northwest National
Laboratory (PNNL) Environmental Molecular Sciences Laboratory (EMSL)
computing group for strong computing support; and the National
Institute of Informatics, and Science Information NETwork 6 (SINET6) for
valuable network support.

\bibliographystyle{apsrev4-2}
\bibliography{paper}


\onecolumngrid
\newpage

\begin{center}
\textbf{Supplemental Material}
\end{center}

\section{Definitions of the Angular Coefficients}
 In the SM the angular coefficients for $B\to D^* \ell \bar{\nu}_\ell$ with $D^* \to D\pi$ can be written in terms of the helicity amplitudes:
 \begin{equation}
 \begin{aligned}
     J_{1s} &= F \left( \frac{1}{2}(H_+^2 + H_-^2) (m_\ell^2 + 3q^2) \right) \,, \\
     J_{1c} &= F \left( 2 (2 m_\ell^2 H_t^2 + H_0^2(m_\ell^2 + q^2))  \right)\,, \\ 
     J_{2s} &= F \left( \frac{1}{2}(H_+^2 + H_-^2) (q^2 - m_\ell^2) \right) \,, \\
     J_{2c} &= F \left( 2 H_0^2 (m_\ell^2 - q^2) \right) \,, \\
     J_{3}  &= F \left( 2 H_+ H_- (m_\ell^2 - q^2)  \right)\,, \\ 
     {J_{4}}  &= -F \left( H_0(H_+ + H_-)(m_\ell^2 - q^2) \right) \,, \\
     J_{5}  &= F \left( -2(H_+ + H_-) H_t m_\ell^2 - 2 H_0 (H_+ - H_-)q^2 \right) \,, \\
     J_{6s} &= -F \left( 2(H_+^2 - H_-^2) q^2 \right) \,, \\
     {J_{6c}} &= -F \left( -8 H_0 H_t m_\ell^2 \right) \,, \\
     J_{7}  &= 0 \,, \\
     J_{8}  &= 0 \,, \\
     J_{9}  &= 0 \,,
 \end{aligned}
 \end{equation}
 with $F = (2 / m_B ^ 3)  (3 p_{D^*}) / (128 \times 2^4  m_{B}^2)$.

\section{Relations Between Common Observables and Angular Coefficients}
The forward-backward asymmetry and the longitudinal polarization fractions are
\begin{align}
    A_{\rm FB} &= \frac{3}{2} \frac{ (J_{6c} + 2 J_{6s}) }{3 J_{1c} - J_{2c} + 2 (3 J_{1s} - J_{2s})} \,, \\
    F_L(D^*) &=  \frac{(3 J_{1c} - J_{2c})}{3 J_{1c} - J_{2c} + 2 (3 J_{1s} - J_{2s})} \,,\\
\end{align}
where the denominator is only a normalization factor.

The $S_i$ observables in the text are directly proportional to the angular coefficients $S_i \propto \hat{J}_i$. The relations are
\begin{align}
    S_3 &= \frac{1}{\pi} \frac{4 J_3}{3 J_{1c} - J_{2c} + 2 (3 J_{1s} - J_{2s})} \,,  \\
    S_5 &= \frac{3 J_5}{3 J_{1c} - J_{2c} + 2 (3 J_{1s} - J_{2s})} \,,  \\
    S_7 &= \frac{3 J_7}{3 J_{1c} - J_{2c} + 2 (3 J_{1s} - J_{2s})} \,,  \\
    S_9 &= \frac{1}{\pi} \frac{4 J_9}{3 J_{1c} - J_{2c} + 2 (3 J_{1s} - J_{2s})} \,,  \\
\end{align}
where the denominator is only a normalization factor. Thus $S_i \propto J_i$. The numerical values are shown in Tab.~\ref{more_observables}. The correlations are provided in the HEPData repository.

\begin{table}[]
    \centering
    \caption{Observables calculated from the average $J_i$ of the four measured decay modes.}
    \begin{tabular}{lrrrrrr}
    \hline
    \hline
    $w$ bin\phantom{00} & $A_{\rm FB}$\phantom{00} & $F_{\rm L}(D^*)$\phantom{00} & $S_3$\phantom{00} & $S_5$\phantom{00} & $S_7$\phantom{00} & $S_9$ \\
    \hline
    $1.00 < w < 1.15$\phantom{00} & $0.23\pm0.03$\phantom{00} & $0.29\pm0.04$\phantom{00} & $-0.11\pm0.02$\phantom{00} & $0.07\pm0.05$\phantom{00} & $0.02\pm0.05  $\phantom{00}&  $0.05\pm0.02$ \\
    $1.15 < w < 1.25$\phantom{00} & $0.30\pm0.03$\phantom{00} & $0.45\pm0.04$\phantom{00} & $-0.06\pm0.02$\phantom{00} & $0.23\pm0.04$\phantom{00} & $0.08\pm0.04  $\phantom{00}&  $0.01\pm0.02$ \\
    $1.25 < w < 1.35$\phantom{00} & $0.29\pm0.03$\phantom{00} & $0.47\pm0.03$\phantom{00} & $-0.03\pm0.02$\phantom{00} & $0.26\pm0.04$\phantom{00} & $-0.06\pm0.04 $\phantom{00} & $ -0.02\pm0.02$ \\
    $1.35 < w < 1.50$\phantom{00} & $0.16\pm0.03$\phantom{00} & $0.70\pm0.03$\phantom{00} & $-0.01\pm0.02$\phantom{00} & $0.20\pm0.03$\phantom{00} & $0.07\pm0.03  $\phantom{00}&  $-0.01\pm0.02$ \\
    \hline
    \hline
    \end{tabular}
    \label{tab:more_observables}
\end{table}

\section{Sytematics}
The uncertainties of the angular coefficients are shown for their individual components in Tab.~\ref{tab:sys15,tab:sys16,tab:sys17,tab:sys18}. The dominant systematic uncertainty is the limited MC statistics.

\begin{table}[]
    \small
    \centering
    \caption{Uncertainties in \% for the $\bar{B}^0\to D^* e \bar{\nu}_e$ channel}
    \input{tableJ_15}
    \label{tab:sys15}
\end{table}

\begin{table}[]
    \small
    \centering
    \caption{Uncertainties in \% for the $\bar{B}^0\to D^* \mu \bar{\nu}_\mu$ channel}
    \input{tableJ_16}
    \label{tab:sys16}
\end{table}

\begin{table}[]
    \small
    \centering
    \caption{Uncertainties in \% for the $\bar{B}^+\to D^* e \bar{\nu}_e$ channel}
    \input{tableJ_17}
    \label{tab:sys17}
\end{table}

\begin{table}[]
    \small
    \centering
    \caption{Uncertainties in \% for the $\bar{B}^+\to D^* \mu \bar{\nu}_\mu$ channel}
    \input{tableJ_18}
    \label{tab:sys18}
\end{table}

\section{$|V_{cb}|$ Fitting Details}
We translate the available beyond zero-recoil lattice QCD prediction for the $B \to D^*$ form factors into the $h_{A_1}$, $R_1$, and $R_2$ representation using the relations
\begin{equation}
    R_1 = \frac{h_V}{h_{A_1}}\, ,\quad
    R_2 = \frac{h_{A_3} + r_{D^*}h_{A_2}}{h_{A_1}}\,
\end{equation}
with $r^* = m_{D^*}/m_B$, and 
\begin{equation}
    \begin{aligned}
         h_{A_1}  = \frac{f}{ m_B \sqrt{r^*} (w+1) } \, , \quad h_V = g \, m_B \, \sqrt{r^*} \, , \\
         h_{A_1} \, (w - r^* - (w-1) \, R_2)  = \frac{F_1}{m_B^2 \, \sqrt{r^*} (w+1)} \, , \\
    \end{aligned}
\end{equation}

The coefficients for the fit described in the main text are shown in Table~\ref{tab:bgl} for the BGL fit, and Table~\ref{tab:cln} for the CLN fit. 
We also investigate the $B^0$ and $B^+$ decay modes independently. We find consistent values for $|V_{cb}|$: 
$|V_{cb}|^{B^0}_{\rm CLN} = 40.9 \pm 0.7$,
$|V_{cb}|^{B^0}_{\rm BGL} = 40.8 \pm 0.7$,
$|V_{cb}|^{B^+}_{\rm CLN} = 41.3 \pm 0.7$,
$|V_{cb}|^{B^+}_{\rm BGL} = 41.3 \pm 0.7$.
Here, we have used the same number of BGL coefficients as the nominal fit.

\begin{table}[]
    \centering
    \caption{The fitted coefficients for the BGL fit described in the text.}
    \label{tab:bgl}
\begin{tabular}{lrrrrrrrrrr}
\hline
\hline
{} &          Value & \multicolumn{9}{l}{Correlation} \\
\hline
$a_0 \times 10^2$      & $    2.68\pm0.06 $&        1.00 &   0.03 &  -0.11 &   0.17 &   0.00 &  -0.02 &   0.08 &  -0.04 &  -0.19 \\
$a_1 \times 10^2$      & $   -2.65\pm2.35 $&        0.03 &   1.00 &  -0.67 &  -0.01 &   0.17 &  -0.03 &   0.09 &  -0.03 &  -0.10 \\
$a_2 \times 10^2$      & $-145.18\pm87.35 $&       -0.11 &  -0.67 &   1.00 &   0.03 &  -0.02 &  -0.03 &   0.01 &  -0.05 &  -0.03 \\
$b_0 \times 10^2$      & $    1.30\pm0.01 $&        0.17 &  -0.01 &   0.03 &   1.00 &  -0.02 &   0.02 &   0.02 &  -0.03 &  -0.50 \\
$b_1 \times 10^2$      & $    1.30\pm0.66 $&        0.00 &   0.17 &  -0.02 &  -0.02 &   1.00 &  -0.66 &   0.59 &  -0.46 &  -0.28 \\
$b_2 \times 10^2$      & $ -11.50\pm22.40 $&       -0.02 &  -0.03 &  -0.03 &   0.02 &  -0.66 &   1.00 &  -0.45 &   0.40 &  -0.04 \\
$c_1 \times 10^2$      & $   -0.25\pm0.16 $&        0.08 &   0.09 &   0.01 &   0.02 &   0.59 &  -0.45 &   1.00 &  -0.82 &  -0.32 \\
$c_2 \times 10^2$      & $    0.78\pm3.29 $&       -0.04 &  -0.03 &  -0.05 &  -0.03 &  -0.46 &   0.40 &  -0.82 &   1.00 &   0.05 \\
$|V_{cb}| \times 10^3$ & $   40.97\pm0.67 $&       -0.19 &  -0.10 &  -0.03 &  -0.50 &  -0.28 &  -0.04 &  -0.32 &   0.05 &   1.00 \\
\hline
\hline
\end{tabular}

\end{table}

\begin{table}[]
    \centering
    \caption{The fitted coefficients for the CLN fit described in the text.}
    \label{tab:cln}
\begin{tabular}{lrrrrrr}
\hline
\hline
{} &         Value & \multicolumn{5}{l}{Correlation} \\
\hline
$h_{A_1}(1)$           & $  0.90\pm0.01 $ &        1.00 &   0.02 &  -0.18 &  -0.04 &  -0.50 \\
$\rho^2$               & $  1.13\pm0.04 $ &        0.02 &   1.00 &   0.05 &  -0.48 &   0.40 \\
$R_1(1)$               & $  1.34\pm0.03 $ &       -0.18 &   0.05 &   1.00 &  -0.10 &  -0.02 \\
$R_2(1)$               & $  1.03\pm0.03 $ &       -0.04 &  -0.48 &  -0.10 &   1.00 &   0.17 \\
$|V_{cb}| \times 10^3$ & $ 40.89\pm0.66 $ &       -0.50 &   0.40 &  -0.02 &   0.17 &   1.00 \\
\hline
\hline
\end{tabular}
\end{table}

\end{document}

%% file: authorsOrcid.tex
\noaffiliation
\author{M.~T.~Prim\,\orcidlink{0000-0002-1407-7450}} 
\author{F.~Bernlochner\,\orcidlink{0000-0001-8153-2719}} 
\author{F.~Metzner\,\orcidlink{0000-0002-0128-264X}} 
  \author{H.~Aihara\,\orcidlink{0000-0002-1907-5964}} 
  \author{D.~M.~Asner\,\orcidlink{0000-0002-1586-5790}} 
  \author{T.~Aushev\,\orcidlink{0000-0002-6347-7055}} 
  \author{R.~Ayad\,\orcidlink{0000-0003-3466-9290}} 
  \author{V.~Babu\,\orcidlink{0000-0003-0419-6912}} 
  \author{Sw.~Banerjee\,\orcidlink{0000-0001-8852-2409}} 
  \author{P.~Behera\,\orcidlink{0000-0002-1527-2266}} 
% ...
\collaboration{The Belle Collaboration}

%% file: tableJ_15.tex
\begin{tabular}{llrrrrrrrrrrr}
\hline
\hline
       &              &  total & $M_\mathrm{miss}^2$ fit & \multicolumn{9}{c}{Unfolding and Acceptance} \\
   &              &   &  & FF($B\to D^*\ell\nu)$ & $\mathcal{B}(D\to X)$ & MC stat. & $\epsilon(\pi_\mathrm{slow})$ & $\epsilon(\mathrm{LID})$ & $\epsilon(\pi^0)$ & $\epsilon$(Tracking) & $\epsilon(K_S^0)$ & FEI Shape \\
$J_i$ & $w$ bin &        &                         &                       &                       &          &                               &                          &                   &                      &                   &           \\
\hline
$J_{1s}$ & [1.00, 1.15) &  23.21 &                   22.32 &                  0.87 &                  0.87 &     6.18 &                          0.84 &                     0.16 &              0.18 &                 0.08 &              0.01 &      0.42 \\
   & [1.15, 1.25) &  14.53 &                   14.02 &                  0.53 &                  0.15 &     3.72 &                          0.23 &                     0.11 &              0.11 &                 0.00 &              0.01 &      0.48 \\
   & [1.25, 1.35) &  11.65 &                   11.17 &                  1.23 &                  0.39 &     3.04 &                          0.29 &                     0.06 &              0.10 &                 0.02 &              0.01 &      0.09 \\
   & [1.35, 1.51) &  10.34 &                    9.97 &                  1.09 &                  0.42 &     2.46 &                          0.22 &                     0.07 &              0.08 &                 0.02 &              0.01 &      0.27 \\
$J_{1c}$ & [1.00, 1.15) &  40.45 &                   38.54 &                  1.01 &                  0.51 &    12.18 &                          0.77 &                     0.20 &              0.00 &                 0.01 &              0.02 &      0.66 \\
   & [1.15, 1.25) &  27.00 &                   26.14 &                  1.26 &                  0.23 &     6.58 &                          0.30 &                     0.09 &              0.17 &                 0.01 &              0.04 &      0.73 \\
   & [1.25, 1.35) &  22.34 &                   21.74 &                  1.20 &                  0.25 &     4.96 &                          0.23 &                     0.23 &              0.01 &                 0.03 &              0.00 &      0.25 \\
   & [1.35, 1.51) &  23.25 &                   22.52 &                  2.23 &                  0.53 &     5.29 &                          0.44 &                     0.47 &              0.08 &                 0.07 &              0.01 &      0.12 \\
$J_{2s}$ & [1.00, 1.15) &  41.57 &                   39.97 &                  0.60 &                  0.66 &    11.38 &                          0.16 &                     0.25 &              0.07 &                 0.04 &              0.02 &      0.34 \\
   & [1.15, 1.25) &  25.08 &                   24.16 &                  1.02 &                  0.34 &     6.62 &                          0.11 &                     0.06 &              0.08 &                 0.03 &              0.02 &      0.11 \\
   & [1.25, 1.35) &  19.69 &                   18.79 &                  1.31 &                  0.29 &     5.72 &                          0.12 &                     0.08 &              0.05 &                 0.01 &              0.01 &      0.36 \\
   & [1.35, 1.51) &  18.96 &                   18.38 &                  1.54 &                  0.14 &     4.40 &                          0.07 &                     0.08 &              0.02 &                 0.01 &              0.01 &      0.14 \\
$J_{2c}$ & [1.00, 1.15) &  70.83 &                   67.77 &                  1.19 &                  0.95 &    20.51 &                          0.35 &                     0.07 &              0.04 &                 0.03 &              0.02 &      1.11 \\
   & [1.15, 1.25) &  46.84 &                   44.80 &                  1.95 &                  0.68 &    13.51 &                          0.24 &                     0.11 &              0.09 &                 0.06 &              0.03 &      0.17 \\
   & [1.25, 1.35) &  39.22 &                   38.15 &                  1.41 &                  0.65 &     8.92 &                          0.35 &                     0.19 &              0.01 &                 0.00 &              0.03 &      0.58 \\
   & [1.35, 1.51) &  45.04 &                   43.74 &                  1.92 &                  0.58 &    10.46 &                          0.44 &                     0.74 &              0.06 &                 0.02 &              0.02 &      1.23 \\
$J_3$ & [1.00, 1.15) &  52.83 &                   51.04 &                  1.10 &                  0.32 &    13.56 &                          0.19 &                     0.09 &              0.00 &                 0.02 &              0.05 &      1.02 \\
   & [1.15, 1.25) &  29.46 &                   28.49 &                  1.15 &                  0.38 &     7.33 &                          0.21 &                     0.07 &              0.06 &                 0.01 &              0.02 &      0.99 \\
   & [1.25, 1.35) &  24.57 &                   23.76 &                  1.47 &                  0.43 &     6.00 &                          0.13 &                     0.15 &              0.11 &                 0.00 &              0.01 &      0.85 \\
   & [1.35, 1.51) &  23.50 &                   22.84 &                  1.77 &                  0.33 &     5.20 &                          0.07 &                     0.04 &              0.05 &                 0.01 &              0.01 &      0.12 \\
$J_4$ & [1.00, 1.15) &  55.54 &                   53.48 &                  0.39 &                  0.49 &    14.93 &                          0.32 &                     0.11 &              0.06 &                 0.00 &              0.02 &      1.18 \\
   & [1.15, 1.25) &  30.94 &                   29.98 &                  0.65 &                  0.39 &     7.58 &                          0.30 &                     0.07 &              0.08 &                 0.00 &              0.01 &      0.88 \\
   & [1.25, 1.35) &  24.83 &                   23.97 &                  0.72 &                  0.35 &     6.44 &                          0.26 &                     0.17 &              0.02 &                 0.04 &              0.01 &      0.22 \\
   & [1.35, 1.51) &  25.86 &                   25.07 &                  1.37 &                  0.71 &     5.96 &                          0.37 &                     0.08 &              0.20 &                 0.03 &              0.01 &      1.48 \\
$J_5$ & [1.00, 1.15) &  43.22 &                   41.89 &                  1.31 &                  0.43 &    10.54 &                          0.43 &                     0.19 &              0.03 &                 0.01 &              0.02 &      0.06 \\
   & [1.15, 1.25) &  25.23 &                   24.38 &                  1.32 &                  0.31 &     6.35 &                          0.12 &                     0.27 &              0.02 &                 0.01 &              0.01 &      0.42 \\
   & [1.25, 1.35) &  20.16 &                   19.32 &                  1.23 &                  0.30 &     5.57 &                          0.12 &                     0.16 &              0.00 &                 0.03 &              0.01 &      0.39 \\
   & [1.35, 1.51) &  18.13 &                   17.56 &                  1.72 &                  0.54 &     4.10 &                          0.32 &                     0.14 &              0.07 &                 0.04 &              0.01 &      0.52 \\
$J_{6s}$ & [1.00, 1.15) &  40.96 &                   39.58 &                  1.76 &                  0.84 &    10.30 &                          0.59 &                     0.28 &              0.22 &                 0.05 &              0.02 &      0.66 \\
   & [1.15, 1.25) &  24.63 &                   23.90 &                  1.31 &                  0.22 &     5.67 &                          0.24 &                     0.47 &              0.08 &                 0.01 &              0.02 &      1.03 \\
   & [1.25, 1.35) &  20.01 &                   19.15 &                  1.38 &                  0.46 &     5.51 &                          0.32 &                     0.54 &              0.16 &                 0.05 &              0.01 &      0.78 \\
   & [1.35, 1.51) &  21.57 &                   20.91 &                  1.44 &                  0.44 &     5.08 &                          0.18 &                     0.50 &              0.05 &                 0.03 &              0.03 &      0.07 \\
$J_{6c}$ & [1.00, 1.15) &  66.26 &                   63.94 &                  0.74 &                  0.52 &    17.36 &                          0.29 &                     0.19 &              0.20 &                 0.00 &              0.02 &      0.37 \\
   & [1.15, 1.25) &  46.82 &                   45.56 &                  0.87 &                  0.85 &    10.59 &                          0.20 &                     0.90 &              0.42 &                 0.03 &              0.03 &      1.34 \\
   & [1.25, 1.35) &  43.08 &                   41.63 &                  1.10 &                  0.38 &    10.86 &                          0.26 &                     1.44 &              0.09 &                 0.04 &              0.02 &      1.18 \\
   & [1.35, 1.51) &  52.87 &                   51.48 &                  1.07 &                  0.37 &    11.69 &                          0.16 &                     2.48 &              0.03 &                 0.02 &              0.04 &      0.47 \\
$J_7$ & [1.00, 1.15) &  43.07 &                   41.77 &                  0.17 &                  0.46 &    10.48 &                          0.09 &                     0.07 &              0.15 &                 0.01 &              0.01 &      0.65 \\
   & [1.15, 1.25) &  24.53 &                   23.81 &                  0.33 &                  0.20 &     5.88 &                          0.12 &                     0.05 &              0.05 &                 0.02 &              0.01 &      0.23 \\
   & [1.25, 1.35) &  19.79 &                   19.21 &                  0.15 &                  0.22 &     4.64 &                          0.09 &                     0.02 &              0.02 &                 0.01 &              0.02 &      1.09 \\
   & [1.35, 1.51) &  17.80 &                   17.31 &                  0.18 &                  0.16 &     4.16 &                          0.08 &                     0.01 &              0.01 &                 0.00 &              0.01 &      0.11 \\
$J_8$ & [1.00, 1.15) &  54.76 &                   52.77 &                  0.17 &                  0.68 &    14.61 &                          0.15 &                     0.03 &              0.06 &                 0.04 &              0.01 &      0.50 \\
   & [1.15, 1.25) &  31.10 &                   29.86 &                  0.18 &                  0.30 &     8.68 &                          0.14 &                     0.06 &              0.07 &                 0.00 &              0.03 &      0.02 \\
   & [1.25, 1.35) &  24.98 &                   23.99 &                  0.24 &                  0.42 &     6.92 &                          0.13 &                     0.09 &              0.06 &                 0.04 &              0.01 &      0.53 \\
   & [1.35, 1.51) &  26.06 &                   25.23 &                  0.31 &                  0.37 &     6.52 &                          0.14 &                     0.05 &              0.09 &                 0.03 &              0.01 &      0.03 \\
$J_9$ & [1.00, 1.15) &  55.59 &                   54.12 &                  0.59 &                  0.84 &    12.64 &                          0.63 &                     0.04 &              0.31 &                 0.00 &              0.05 &      0.41 \\
   & [1.15, 1.25) &  30.19 &                   29.00 &                  0.22 &                  0.26 &     8.38 &                          0.10 &                     0.02 &              0.01 &                 0.00 &              0.01 &      0.33 \\
   & [1.25, 1.35) &  25.00 &                   24.22 &                  0.10 &                  0.23 &     6.17 &                          0.10 &                     0.01 &              0.09 &                 0.01 &              0.02 &      0.31 \\
   & [1.35, 1.51) &  23.36 &                   22.94 &                  0.12 &                  0.21 &     4.37 &                          0.05 &                     0.06 &              0.03 &                 0.02 &              0.01 &      0.24 \\
\hline
\hline
\end{tabular}

%% file: tableJ_16.tex
\begin{tabular}{llrrrrrrrrrrr}
\hline
\hline
       &              &  total & $M_\mathrm{miss}^2$ fit & \multicolumn{9}{c}{Unfolding and Acceptance} \\
   &              &   &  & FF($B\to D^*\ell\nu)$ & $\mathcal{B}(D\to X)$ & MC stat. & $\epsilon(\pi_\mathrm{slow})$ & $\epsilon(\mathrm{LID})$ & $\epsilon(\pi^0)$ & $\epsilon$(Tracking) & $\epsilon(K_S^0)$ & FEI Shape \\
$J_i$ & $w$ bin &        &                         &                       &                       &          &                               &                          &                   &                      &                   &           \\
\hline
$J_{1s}$ & [1.00, 1.15) &  23.53 &                   22.42 &                  1.19 &                  1.32 &     6.82 &                          0.99 &                     0.05 &              0.28 &                 0.12 &              0.03 &      0.57 \\
   & [1.15, 1.25) &  13.42 &                   12.81 &                  0.62 &                  0.13 &     3.93 &                          0.20 &                     0.05 &              0.01 &                 0.01 &              0.01 &      0.02 \\
   & [1.25, 1.35) &  10.89 &                   10.38 &                  1.17 &                  0.43 &     3.04 &                          0.29 &                     0.14 &              0.10 &                 0.04 &              0.01 &      0.00 \\
   & [1.35, 1.51) &  10.26 &                    9.85 &                  1.19 &                  0.20 &     2.56 &                          0.20 &                     0.09 &              0.02 &                 0.02 &              0.01 &      0.22 \\
$J_{1c}$ & [1.00, 1.15) &  35.81 &                   34.39 &                  0.84 &                  0.30 &     9.92 &                          0.90 &                     0.23 &              0.06 &                 0.01 &              0.01 &      0.16 \\
   & [1.15, 1.25) &  20.48 &                   19.63 &                  1.06 &                  0.26 &     5.72 &                          0.22 &                     0.02 &              0.05 &                 0.01 &              0.01 &      0.19 \\
   & [1.25, 1.35) &  20.05 &                   19.27 &                  1.06 &                  0.24 &     5.41 &                          0.20 &                     0.12 &              0.07 &                 0.03 &              0.00 &      0.03 \\
   & [1.35, 1.51) &  28.93 &                   27.76 &                  2.40 &                  1.16 &     7.64 &                          0.65 &                     0.25 &              0.40 &                 0.08 &              0.03 &      0.07 \\
$J_{2s}$ & [1.00, 1.15) &  43.84 &                   42.06 &                  0.64 &                  0.60 &    12.32 &                          0.63 &                     0.18 &              0.03 &                 0.07 &              0.03 &      0.22 \\
   & [1.15, 1.25) &  23.59 &                   22.61 &                  0.96 &                  0.20 &     6.62 &                          0.11 &                     0.11 &              0.02 &                 0.03 &              0.01 &      0.24 \\
   & [1.25, 1.35) &  18.82 &                   17.91 &                  1.06 &                  0.23 &     5.65 &                          0.14 &                     0.19 &              0.03 &                 0.02 &              0.01 &      0.14 \\
   & [1.35, 1.51) &  19.83 &                   19.08 &                  1.74 &                  0.17 &     5.13 &                          0.11 &                     0.04 &              0.07 &                 0.02 &              0.01 &      0.12 \\
$J_{2c}$ & [1.00, 1.15) &  62.30 &                   60.17 &                  0.55 &                  0.96 &    16.08 &                          0.65 &                     0.07 &              0.07 &                 0.09 &              0.03 &      0.46 \\
   & [1.15, 1.25) &  37.32 &                   36.30 &                  1.84 &                  0.27 &     8.49 &                          0.21 &                     0.27 &              0.07 &                 0.07 &              0.03 &      0.20 \\
   & [1.25, 1.35) &  36.58 &                   35.48 &                  1.18 &                  0.45 &     8.81 &                          0.36 &                     0.11 &              0.06 &                 0.07 &              0.02 &      0.18 \\
   & [1.35, 1.51) &  55.71 &                   53.90 &                  2.11 &                  0.51 &    13.78 &                          0.29 &                     0.14 &              0.16 &                 0.05 &              0.02 &      1.91 \\
$J_{3}$ & [1.00, 1.15) &  49.51 &                   47.65 &                  1.20 &                  1.12 &    13.25 &                          0.42 &                     0.15 &              0.45 &                 0.03 &              0.02 &      1.49 \\
   & [1.15, 1.25) &  29.74 &                   28.80 &                  1.07 &                  0.37 &     7.30 &                          0.20 &                     0.04 &              0.06 &                 0.05 &              0.02 &      0.66 \\
   & [1.25, 1.35) &  23.65 &                   22.81 &                  1.52 &                  0.36 &     5.99 &                          0.20 &                     0.03 &              0.05 &                 0.03 &              0.01 &      0.64 \\
   & [1.35, 1.51) &  25.05 &                   24.28 &                  1.63 &                  0.26 &     5.94 &                          0.16 &                     0.10 &              0.04 &                 0.00 &              0.01 &      0.22 \\
$J_{4}$ & [1.00, 1.15) &  48.88 &                   47.60 &                  0.45 &                  0.72 &    10.88 &                          0.49 &                     0.28 &              0.17 &                 0.01 &              0.02 &      2.03 \\
   & [1.15, 1.25) &  29.96 &                   28.90 &                  0.98 &                  0.53 &     7.82 &                          0.26 &                     0.10 &              0.15 &                 0.03 &              0.01 &      0.07 \\
   & [1.25, 1.35) &  24.76 &                   23.66 &                  0.65 &                  0.48 &     7.21 &                          0.31 &                     0.20 &              0.12 &                 0.06 &              0.02 &      0.56 \\
   & [1.35, 1.51) &  28.43 &                   27.52 &                  1.09 &                  0.59 &     6.99 &                          0.35 &                     0.20 &              0.17 &                 0.04 &              0.01 &      0.60 \\
$J_{5}$ & [1.00, 1.15) &  40.88 &                   39.50 &                  1.39 &                  0.53 &    10.41 &                          0.37 &                     0.06 &              0.01 &                 0.01 &              0.03 &      0.62 \\
   & [1.15, 1.25) &  24.57 &                   23.45 &                  1.58 &                  0.44 &     7.08 &                          0.14 &                     0.18 &              0.05 &                 0.04 &              0.02 &      0.95 \\
   & [1.25, 1.35) &  19.77 &                   18.83 &                  1.37 &                  0.34 &     5.84 &                          0.36 &                     0.08 &              0.09 &                 0.03 &              0.02 &      0.39 \\
   & [1.35, 1.51) &  19.41 &                   18.88 &                  1.27 &                  0.29 &     4.30 &                          0.23 &                     0.08 &              0.02 &                 0.01 &              0.01 &      0.13 \\
$J_{6s}$ & [1.00, 1.15) &  39.72 &                   38.19 &                  1.78 &                  1.20 &    10.62 &                          1.10 &                     0.18 &              0.31 &                 0.10 &              0.02 &      0.58 \\
   & [1.15, 1.25) &  25.38 &                   24.34 &                  1.18 &                  0.28 &     7.07 &                          0.21 &                     0.03 &              0.02 &                 0.00 &              0.01 &      0.66 \\
   & [1.25, 1.35) &  20.07 &                   19.35 &                  1.28 &                  0.44 &     5.12 &                          0.31 &                     0.22 &              0.08 &                 0.05 &              0.02 &      0.19 \\
   & [1.35, 1.51) &  24.35 &                   23.79 &                  1.60 &                  0.64 &     4.90 &                          0.20 &                     0.14 &              0.07 &                 0.03 &              0.00 &      0.02 \\
$J_{6c}$ & [1.00, 1.15) &  65.84 &                   63.82 &                  0.73 &                  0.50 &    16.13 &                          0.48 &                     0.04 &              0.25 &                 0.00 &              0.02 &      0.13 \\
   & [1.15, 1.25) &  43.21 &                   41.57 &                  1.00 &                  0.41 &    11.69 &                          0.19 &                     0.09 &              0.12 &                 0.01 &              0.01 &      1.27 \\
   & [1.25, 1.35) &  41.82 &                   40.79 &                  1.12 &                  0.21 &     9.12 &                          0.18 &                     0.07 &              0.02 &                 0.06 &              0.02 &      0.43 \\
   & [1.35, 1.51) &  61.00 &                   59.72 &                  1.29 &                  0.80 &    12.24 &                          0.22 &                     0.12 &              0.07 &                 0.01 &              0.02 &      1.61 \\
$J_{7}$ & [1.00, 1.15) &  40.48 &                   39.05 &                  0.15 &                  0.36 &    10.66 &                          0.20 &                     0.09 &              0.02 &                 0.02 &              0.03 &      0.63 \\
   & [1.15, 1.25) &  23.87 &                   23.12 &                  0.20 &                  0.18 &     5.81 &                          0.12 &                     0.11 &              0.03 &                 0.03 &              0.01 &      1.26 \\
   & [1.25, 1.35) &  18.96 &                   18.22 &                  0.28 &                  0.27 &     5.19 &                          0.13 &                     0.02 &              0.02 &                 0.01 &              0.00 &      0.82 \\
   & [1.35, 1.51) &  19.06 &                   18.51 &                  0.13 &                  0.10 &     4.52 &                          0.07 &                     0.05 &              0.04 &                 0.00 &              0.01 &      0.39 \\
$J_{8}$ & [1.00, 1.15) &  50.17 &                   48.05 &                  0.26 &                  0.50 &    14.37 &                          0.53 &                     0.08 &              0.19 &                 0.01 &              0.02 &      1.25 \\
   & [1.15, 1.25) &  29.21 &                   28.15 &                  0.51 &                  0.24 &     7.71 &                          0.17 &                     0.03 &              0.00 &                 0.02 &              0.01 &      1.04 \\
   & [1.25, 1.35) &  23.51 &                   22.59 &                  0.23 &                  0.31 &     6.41 &                          0.10 &                     0.02 &              0.12 &                 0.00 &              0.01 &      1.09 \\
   & [1.35, 1.51) &  27.80 &                   26.94 &                  0.06 &                  0.20 &     6.85 &                          0.05 &                     0.14 &              0.00 &                 0.00 &              0.01 &      0.28 \\
$J_{9}$ & [1.00, 1.15) &  51.08 &                   49.14 &                  0.25 &                  0.64 &    13.73 &                          0.53 &                     0.03 &              0.01 &                 0.07 &              0.09 &      2.10 \\
   & [1.15, 1.25) &  29.01 &                   28.04 &                  0.58 &                  0.33 &     7.36 &                          0.16 &                     0.03 &              0.11 &                 0.01 &              0.02 &      0.60 \\
   & [1.25, 1.35) &  23.92 &                   23.13 &                  0.23 &                  0.34 &     6.05 &                          0.13 &                     0.06 &              0.14 &                 0.02 &              0.02 &      0.29 \\
   & [1.35, 1.51) &  24.85 &                   24.09 &                  0.04 &                  0.25 &     6.10 &                          0.11 &                     0.02 &              0.00 &                 0.02 &              0.01 &      0.15 \\
\hline
\hline
\end{tabular}

%% file: tableJ_17.tex
\begin{tabular}{llrrrrrrrrrrr}
\hline
\hline
       &              &  total & $M_\mathrm{miss}^2$ fit & \multicolumn{9}{c}{Unfolding and Acceptance} \\
   &              &   &  & FF($B\to D^*\ell\nu)$ & $\mathcal{B}(D\to X)$ & MC stat. & $\epsilon(\pi_\mathrm{slow})$ & $\epsilon(\mathrm{LID})$ & $\epsilon(\pi^0)$ & $\epsilon$(Tracking) & $\epsilon(K_S^0)$ & FEI Shape \\
$J_i$ & $w$ bin &        &                         &                       &                       &          &                               &                          &                   &                      &                   &           \\
\hline
$J_{1s}$ & [1.00, 1.15) &  15.39 &                   14.79 &                  1.07 &                  0.24 &     4.08 &                          0.40 &                     0.20 &              0.01 &                 0.01 &              0.01 &      0.43 \\
   & [1.15, 1.25) &  12.77 &                   12.35 &                  0.56 &                  0.15 &     3.17 &                          0.44 &                     0.04 &              0.01 &                 0.01 &              0.01 &      0.01 \\
   & [1.25, 1.35) &  11.83 &                   11.41 &                  1.13 &                  0.11 &     2.89 &                          0.43 &                     0.07 &              0.03 &                 0.00 &              0.00 &      0.10 \\
   & [1.35, 1.51) &  13.19 &                   12.88 &                  0.96 &                  0.15 &     2.62 &                          0.35 &                     0.07 &              0.01 &                 0.01 &              0.01 &      0.40 \\
$J_{1c}$ & [1.00, 1.15) &  29.25 &                   28.28 &                  0.91 &                  0.21 &     7.20 &                          1.00 &                     0.17 &              0.04 &                 0.00 &              0.01 &      1.34 \\
   & [1.15, 1.25) &  23.67 &                   22.88 &                  1.05 &                  0.18 &     5.89 &                          0.75 &                     0.09 &              0.06 &                 0.01 &              0.01 &      0.09 \\
   & [1.25, 1.35) &  19.28 &                   18.82 &                  1.03 &                  0.18 &     3.99 &                          0.42 &                     0.22 &              0.06 &                 0.01 &              0.01 &      0.32 \\
   & [1.35, 1.51) &  30.08 &                   29.52 &                  1.96 &                  0.26 &     5.35 &                          0.29 &                     0.52 &              0.03 &                 0.00 &              0.01 &      0.50 \\
$J_{2s}$ & [1.00, 1.15) &  25.89 &                   25.01 &                  0.41 &                  0.36 &     6.65 &                          0.13 &                     0.15 &              0.04 &                 0.00 &              0.00 &      0.65 \\
   & [1.15, 1.25) &  22.28 &                   21.59 &                  0.98 &                  0.42 &     5.39 &                          0.11 &                     0.09 &              0.03 &                 0.01 &              0.00 &      0.02 \\
   & [1.25, 1.35) &  20.89 &                   20.10 &                  1.27 &                  0.25 &     5.52 &                          0.11 &                     0.05 &              0.02 &                 0.01 &              0.01 &      0.02 \\
   & [1.35, 1.51) &  25.60 &                   25.06 &                  1.33 &                  0.29 &     4.95 &                          0.15 &                     0.11 &              0.00 &                 0.01 &              0.01 &      0.97 \\
$J_{2c}$ & [1.00, 1.15) &  48.37 &                   46.80 &                  1.02 &                  0.65 &    12.09 &                          0.37 &                     0.13 &              0.09 &                 0.02 &              0.02 &      1.36 \\
   & [1.15, 1.25) &  41.41 &                   40.49 &                  1.61 &                  0.64 &     8.49 &                          0.22 &                     0.25 &              0.09 &                 0.02 &              0.01 &      0.07 \\
   & [1.25, 1.35) &  36.08 &                   34.98 &                  1.27 &                  0.38 &     8.73 &                          0.27 &                     0.15 &              0.04 &                 0.01 &              0.01 &      0.57 \\
   & [1.35, 1.51) &  60.89 &                   59.81 &                  2.49 &                  0.68 &    10.97 &                          0.38 &                     0.73 &              0.01 &                 0.06 &              0.02 &      1.68 \\
$J_{3}$ & [1.00, 1.15) &  34.53 &                   33.69 &                  1.14 &                  0.40 &     7.45 &                          0.26 &                     0.27 &              0.00 &                 0.02 &              0.01 &      0.53 \\
   & [1.15, 1.25) &  30.71 &                   29.92 &                  1.18 &                  0.23 &     6.85 &                          0.15 &                     0.04 &              0.08 &                 0.06 &              0.02 &      0.12 \\
   & [1.25, 1.35) &  28.58 &                   27.75 &                  1.39 &                  0.22 &     6.69 &                          0.08 &                     0.07 &              0.15 &                 0.04 &              0.00 &      0.16 \\
   & [1.35, 1.51) &  33.15 &                   32.56 &                  1.63 &                  0.21 &     6.01 &                          0.11 &                     0.03 &              0.01 &                 0.02 &              0.02 &      0.33 \\
$J_{4}$ & [1.00, 1.15) &  33.48 &                   32.66 &                  0.46 &                  0.40 &     7.16 &                          0.48 &                     0.14 &              0.08 &                 0.04 &              0.03 &      1.54 \\
   & [1.15, 1.25) &  30.59 &                   29.82 &                  0.68 &                  0.58 &     6.72 &                          0.23 &                     0.03 &              0.09 &                 0.06 &              0.01 &      0.19 \\
   & [1.25, 1.35) &  27.63 &                   26.85 &                  0.74 &                  0.46 &     6.44 &                          0.22 &                     0.07 &              0.03 &                 0.00 &              0.01 &      0.80 \\
   & [1.35, 1.51) &  36.23 &                   35.39 &                  1.24 &                  0.49 &     7.54 &                          0.12 &                     0.39 &              0.08 &                 0.01 &              0.00 &      1.00 \\
$J_{5}$ & [1.00, 1.15) &  27.28 &                   26.47 &                  1.43 &                  0.13 &     6.43 &                          0.21 &                     0.12 &              0.00 &                 0.02 &              0.01 &      0.12 \\
   & [1.15, 1.25) &  25.01 &                   24.29 &                  1.31 &                  0.08 &     5.81 &                          0.09 &                     0.34 &              0.04 &                 0.02 &              0.00 &      0.20 \\
   & [1.25, 1.35) &  22.30 &                   21.63 &                  1.21 &                  0.31 &     5.27 &                          0.17 &                     0.20 &              0.03 &                 0.01 &              0.00 &      0.18 \\
   & [1.35, 1.51) &  25.54 &                   24.93 &                  1.39 &                  0.23 &     5.36 &                          0.09 &                     0.26 &              0.02 &                 0.00 &              0.01 &      0.09 \\
$J_{6s}$ & [1.00, 1.15) &  25.91 &                   25.01 &                  2.14 &                  0.27 &     6.39 &                          0.26 &                     0.32 &              0.12 &                 0.03 &              0.01 &      0.44 \\
   & [1.15, 1.25) &  23.67 &                   22.81 &                  1.27 &                  0.17 &     6.16 &                          0.39 &                     0.58 &              0.01 &                 0.02 &              0.01 &      0.42 \\
   & [1.25, 1.35) &  21.67 &                   21.12 &                  1.17 &                  0.16 &     4.68 &                          0.35 &                     0.52 &              0.01 &                 0.01 &              0.01 &      0.21 \\
   & [1.35, 1.51) &  28.27 &                   27.62 &                  1.34 &                  0.35 &     5.79 &                          0.19 &                     0.58 &              0.00 &                 0.04 &              0.02 &      0.49 \\
$J_{6c}$ & [1.00, 1.15) &  49.20 &                   47.55 &                  0.92 &                  0.49 &    12.57 &                          0.79 &                     0.19 &              0.19 &                 0.05 &              0.01 &      0.60 \\
   & [1.15, 1.25) &  44.76 &                   43.46 &                  0.82 &                  0.16 &    10.66 &                          0.56 &                     0.49 &              0.03 &                 0.01 &              0.02 &      0.19 \\
   & [1.25, 1.35) &  41.60 &                   40.50 &                  1.11 &                  0.27 &     9.34 &                          0.34 &                     1.11 &              0.09 &                 0.02 &              0.01 &      0.21 \\
   & [1.35, 1.51) &  63.43 &                   62.16 &                  1.32 &                  0.75 &    12.34 &                          0.17 &                     2.21 &              0.05 &                 0.08 &              0.02 &      0.37 \\
$J_{7}$ & [1.00, 1.15) &  27.21 &                   26.48 &                  0.19 &                  0.17 &     6.26 &                          0.17 &                     0.14 &              0.05 &                 0.05 &              0.02 &      0.33 \\
   & [1.15, 1.25) &  24.31 &                   23.60 &                  0.28 &                  0.39 &     5.80 &                          0.11 &                     0.08 &              0.05 &                 0.03 &              0.01 &      0.21 \\
   & [1.25, 1.35) &  21.71 &                   21.23 &                  0.26 &                  0.17 &     4.50 &                          0.16 &                     0.15 &              0.06 &                 0.01 &              0.01 &      0.40 \\
   & [1.35, 1.51) &  25.07 &                   24.56 &                  0.53 &                  0.16 &     4.97 &                          0.16 &                     0.15 &              0.04 &                 0.00 &              0.02 &      0.17 \\
$J_{8}$ & [1.00, 1.15) &  34.28 &                   33.26 &                  0.20 &                  0.43 &     8.21 &                          0.19 &                     0.05 &              0.09 &                 0.03 &              0.03 &      0.85 \\
   & [1.15, 1.25) &  30.07 &                   29.33 &                  0.31 &                  0.26 &     6.55 &                          0.14 &                     0.04 &              0.05 &                 0.02 &              0.01 &      1.18 \\
   & [1.25, 1.35) &  27.11 &                   26.49 &                  0.08 &                  0.44 &     5.71 &                          0.20 &                     0.11 &              0.00 &                 0.02 &              0.01 &      0.61 \\
   & [1.35, 1.51) &  35.30 &                   34.60 &                  0.12 &                  0.42 &     6.95 &                          0.18 &                     0.12 &              0.02 &                 0.03 &              0.02 &      0.57 \\
$J_{9}$ & [1.00, 1.15) &  34.17 &                   33.15 &                  0.38 &                  0.36 &     8.24 &                          0.32 &                     0.13 &              0.02 &                 0.05 &              0.04 &      0.09 \\
   & [1.15, 1.25) &  30.88 &                   30.02 &                  0.14 &                  0.14 &     7.22 &                          0.19 &                     0.03 &              0.06 &                 0.01 &              0.01 &      0.65 \\
   & [1.25, 1.35) &  27.15 &                   26.61 &                  0.17 &                  0.26 &     5.36 &                          0.16 &                     0.14 &              0.01 &                 0.00 &              0.01 &      0.11 \\
   & [1.35, 1.51) &  32.87 &                   32.29 &                  0.29 &                  0.27 &     6.14 &                          0.11 &                     0.01 &              0.00 &                 0.01 &              0.02 &      0.31 \\
\hline
\hline
\end{tabular}

%% file: tableJ_18.tex
\begin{tabular}{llrrrrrrrrrrr}
\hline
\hline
       &              &  total & $M_\mathrm{miss}^2$ fit & \multicolumn{9}{c}{Unfolding and Acceptance} \\
   &              &   &  & FF($B\to D^*\ell\nu)$ & $\mathcal{B}(D\to X)$ & MC stat. & $\epsilon(\pi_\mathrm{slow})$ & $\epsilon(\mathrm{LID})$ & $\epsilon(\pi^0)$ & $\epsilon$(Tracking) & $\epsilon(K_S^0)$ & FEI Shape \\
$J_i$ & $w$ bin &        &                         &                       &                       &          &                               &                          &                   &                      &                   &           \\
\hline
$J_{1s}$ & [1.00, 1.15) &  12.91 &                   12.21 &                  1.18 &                  0.19 &     4.01 &                          0.44 &                     0.06 &              0.06 &                 0.00 &              0.02 &      0.18 \\
   & [1.15, 1.25) &  12.93 &                   12.42 &                  0.62 &                  0.15 &     3.49 &                          0.45 &                     0.02 &              0.01 &                 0.02 &              0.01 &      0.05 \\
   & [1.25, 1.35) &  11.06 &                   10.67 &                  0.81 &                  0.12 &     2.80 &                          0.35 &                     0.06 &              0.02 &                 0.00 &              0.01 &      0.06 \\
   & [1.35, 1.51) &  13.91 &                   13.48 &                  1.07 &                  0.32 &     3.15 &                          0.35 &                     0.02 &              0.05 &                 0.01 &              0.01 &      0.72 \\
$J_{1c}$ & [1.00, 1.15) &  23.29 &                   22.32 &                  0.99 &                  0.37 &     6.51 &                          0.89 &                     0.05 &              0.09 &                 0.02 &              0.01 &      0.04 \\
   & [1.15, 1.25) &  25.39 &                   24.29 &                  1.33 &                  0.42 &     7.25 &                          0.66 &                     0.02 &              0.00 &                 0.06 &              0.01 &      0.00 \\
   & [1.25, 1.35) &  21.98 &                   21.18 &                  0.92 &                  0.14 &     5.75 &                          0.44 &                     0.14 &              0.02 &                 0.01 &              0.01 &      0.18 \\
   & [1.35, 1.51) &  34.53 &                   33.70 &                  2.06 &                  0.99 &     7.07 &                          0.38 &                     0.11 &              0.13 &                 0.05 &              0.02 &      1.09 \\
$J_{2s}$ & [1.00, 1.15) &  22.07 &                   21.08 &                  0.42 &                  0.38 &     6.50 &                          0.31 &                     0.05 &              0.10 &                 0.02 &              0.03 &      0.18 \\
   & [1.15, 1.25) &  21.25 &                   20.38 &                  1.00 &                  0.19 &     5.92 &                          0.18 &                     0.05 &              0.00 &                 0.00 &              0.02 &      0.31 \\
   & [1.25, 1.35) &  19.44 &                   18.85 &                  0.76 &                  0.27 &     4.70 &                          0.08 &                     0.07 &              0.02 &                 0.01 &              0.01 &      0.08 \\
   & [1.35, 1.51) &  25.89 &                   25.05 &                  1.50 &                  0.48 &     6.31 &                          0.11 &                     0.02 &              0.02 &                 0.00 &              0.02 &      0.70 \\
$J_{2c}$ & [1.00, 1.15) &  39.79 &                   38.41 &                  1.04 &                  0.81 &    10.29 &                          0.60 &                     0.17 &              0.06 &                 0.02 &              0.04 &      0.06 \\
   & [1.15, 1.25) &  41.77 &                   39.77 &                  2.01 &                  0.34 &    12.60 &                          0.16 &                     0.20 &              0.03 &                 0.04 &              0.04 &      0.18 \\
   & [1.25, 1.35) &  39.00 &                   37.80 &                  1.17 &                  0.44 &     9.53 &                          0.18 &                     0.19 &              0.06 &                 0.03 &              0.02 &      0.49 \\
   & [1.35, 1.51) &  64.25 &                   62.65 &                  2.16 &                  1.30 &    13.89 &                          0.35 &                     0.38 &              0.11 &                 0.04 &              0.03 &      1.94 \\
$J_{3}$ & [1.00, 1.15) &  29.51 &                   28.33 &                  1.32 &                  0.35 &     8.08 &                          0.30 &                     0.11 &              0.08 &                 0.06 &              0.01 &      0.99 \\
   & [1.15, 1.25) &  28.99 &                   28.05 &                  1.35 &                  0.43 &     7.18 &                          0.17 &                     0.11 &              0.17 &                 0.00 &              0.02 &      0.32 \\
   & [1.25, 1.35) &  27.11 &                   26.23 &                  1.30 &                  0.27 &     6.63 &                          0.16 &                     0.03 &              0.06 &                 0.01 &              0.01 &      1.13 \\
   & [1.35, 1.51) &  33.10 &                   32.39 &                  1.37 &                  0.43 &     6.68 &                          0.13 &                     0.02 &              0.03 &                 0.00 &              0.02 &      0.38 \\
$J_{4}$ & [1.00, 1.15) &  31.48 &                   30.24 &                  0.43 &                  0.48 &     8.69 &                          0.30 &                     0.05 &              0.10 &                 0.02 &              0.05 &      0.43 \\
   & [1.15, 1.25) &  29.33 &                   28.26 &                  1.06 &                  0.44 &     7.75 &                          0.34 &                     0.03 &              0.04 &                 0.03 &              0.02 &      0.06 \\
   & [1.25, 1.35) &  27.09 &                   26.30 &                  0.12 &                  0.21 &     6.45 &                          0.15 &                     0.07 &              0.00 &                 0.02 &              0.01 &      0.76 \\
   & [1.35, 1.51) &  37.61 &                   36.63 &                  0.93 &                  0.37 &     8.46 &                          0.21 &                     0.10 &              0.00 &                 0.01 &              0.02 &      0.25 \\
$J_{5}$ & [1.00, 1.15) &  25.19 &                   24.01 &                  1.43 &                  0.46 &     7.45 &                          0.18 &                     0.08 &              0.10 &                 0.00 &              0.03 &      0.40 \\
   & [1.15, 1.25) &  23.80 &                   22.91 &                  1.71 &                  0.23 &     6.19 &                          0.19 &                     0.21 &              0.10 &                 0.00 &              0.01 &      0.12 \\
   & [1.25, 1.35) &  22.01 &                   21.29 &                  0.95 &                  0.23 &     5.47 &                          0.12 &                     0.02 &              0.04 &                 0.00 &              0.01 &      0.48 \\
   & [1.35, 1.51) &  27.09 &                   26.26 &                  1.03 &                  0.20 &     6.57 &                          0.21 &                     0.06 &              0.03 &                 0.02 &              0.01 &      0.34 \\
$J_{6s}$ & [1.00, 1.15) &  22.73 &                   21.76 &                  2.08 &                  0.53 &     6.18 &                          0.27 &                     0.19 &              0.02 &                 0.03 &              0.03 &      0.33 \\
   & [1.15, 1.25) &  21.92 &                   21.04 &                  1.28 &                  0.39 &     5.99 &                          0.44 &                     0.06 &              0.06 &                 0.01 &              0.02 &      0.14 \\
   & [1.25, 1.35) &  22.39 &                   21.89 &                  1.04 &                  0.09 &     4.59 &                          0.35 &                     0.07 &              0.01 &                 0.00 &              0.01 &      0.07 \\
   & [1.35, 1.51) &  29.98 &                   29.42 &                  1.42 &                  0.19 &     5.58 &                          0.27 &                     0.02 &              0.07 &                 0.01 &              0.01 &      0.21 \\
$J_{6c}$ & [1.00, 1.15) &  42.62 &                   41.23 &                  0.68 &                  0.97 &    10.71 &                          0.55 &                     0.05 &              0.02 &                 0.06 &              0.02 &      0.69 \\
   & [1.15, 1.25) &  43.40 &                   41.88 &                  0.84 &                  0.78 &    11.26 &                          0.56 &                     0.18 &              0.12 &                 0.01 &              0.03 &      0.99 \\
   & [1.25, 1.35) &  46.03 &                   45.08 &                  0.93 &                  0.26 &     9.26 &                          0.50 &                     0.09 &              0.13 &                 0.01 &              0.03 &      0.42 \\
   & [1.35, 1.51) &  72.42 &                   71.07 &                  1.20 &                  0.35 &    13.86 &                          0.34 &                     0.12 &              0.08 &                 0.00 &              0.04 &      0.45 \\
$J_{7}$ & [1.00, 1.15) &  23.68 &                   22.83 &                  0.10 &                  0.36 &     6.26 &                          0.19 &                     0.09 &              0.13 &                 0.02 &              0.01 &      0.55 \\
   & [1.15, 1.25) &  22.11 &                   21.44 &                  0.27 &                  0.13 &     5.31 &                          0.11 &                     0.02 &              0.01 &                 0.04 &              0.02 &      0.76 \\
   & [1.25, 1.35) &  21.72 &                   20.84 &                  0.06 &                  0.13 &     6.12 &                          0.15 &                     0.06 &              0.02 &                 0.00 &              0.00 &      0.01 \\
   & [1.35, 1.51) &  25.41 &                   24.70 &                  0.46 &                  0.31 &     5.93 &                          0.16 &                     0.02 &              0.03 &                 0.03 &              0.01 &      0.06 \\
$J_{8}$ & [1.00, 1.15) &  29.02 &                   28.33 &                  0.18 &                  0.34 &     6.26 &                          0.22 &                     0.09 &              0.00 &                 0.04 &              0.01 &      0.35 \\
   & [1.15, 1.25) &  27.20 &                   26.15 &                  0.41 &                  0.26 &     7.41 &                          0.11 &                     0.04 &              0.06 &                 0.04 &              0.00 &      0.94 \\
   & [1.25, 1.35) &  27.14 &                   26.35 &                  0.14 &                  0.19 &     6.50 &                          0.13 &                     0.08 &              0.02 &                 0.03 &              0.01 &      0.31 \\
   & [1.35, 1.51) &  36.51 &                   35.61 &                  0.32 &                  0.54 &     8.04 &                          0.19 &                     0.03 &              0.01 &                 0.03 &              0.01 &      0.40 \\
$J_{9}$ & [1.00, 1.15) &  30.06 &                   29.09 &                  0.19 &                  0.77 &     7.55 &                          0.19 &                     0.10 &              0.04 &                 0.09 &              0.01 &      0.15 \\
   & [1.15, 1.25) &  28.55 &                   27.41 &                  0.27 &                  0.20 &     7.94 &                          0.20 &                     0.05 &              0.09 &                 0.01 &              0.02 &      0.75 \\
   & [1.25, 1.35) &  27.68 &                   26.80 &                  0.07 &                  0.27 &     6.92 &                          0.09 &                     0.01 &              0.00 &                 0.02 &              0.01 &      0.32 \\
   & [1.35, 1.51) &  32.80 &                   32.17 &                  0.14 &                  0.26 &     6.40 &                          0.11 &                     0.05 &              0.04 &                 0.02 &              0.01 &      0.29 \\
\hline
\hline
\end{tabular}